\RequirePackage{lineno}
\documentclass[twocolumn,prl,superscriptaddress,amsmath,amssymb]{revtex4}

\usepackage{graphicx}% Include figure files
\usepackage{dcolumn}% Align table columns on decimal point
\usepackage{bm}% bold math

\usepackage{xcolor}
\usepackage{lipsum}

\usepackage{graphicx,amsmath,amssymb}
\usepackage{epstopdf}

\begin{document}

%\linenumbers

\title{Measurement of the $Q^2$ dependence of the Deuteron Spin Structure Function $g_1$ and its Moments at Low $Q^2$ with CLAS}

\author{K.P. Adhikari}
\affiliation{Old Dominion University, Norfolk, Virginia 23529, USA }
\affiliation{Thomas Jefferson National Accelerator Facility, Newport News, Virginia 23606, USA }
\affiliation{Mississippi State University, Mississippi State, Mississippi, MS 39762-5167, USA }
\author{A. Deur \footnote{Contact author. Email: deurpam@jlab.org}}
\affiliation{Thomas Jefferson National Accelerator Facility, Newport News, Virginia 23606, USA }
\affiliation{University of Virginia, Charlottesville, Virginia 22901, USA}
\author{L. El Fassi}
\affiliation{Old Dominion University, Norfolk, Virginia 23529, USA}
\affiliation{Mississippi State University, Mississippi State, Mississippi, MS 39762-5167, USA}
\author{H. Kang}
\affiliation{Seoul National University, Seoul, Korea}
\author{S.E. Kuhn} 
\affiliation{Old Dominion University, Norfolk, Virginia 23529, USA}
\author{M. Ripani}
\affiliation{INFN, Sezione di Genova, 16146 Genova, Italy}
\author{K. Slifer}
\affiliation{University of Virginia, Charlottesville, Virginia 22901, USA}
\affiliation{University of New Hampshire, Durham, New Hampshire 03824-3568, USA}
\author{X. Zheng}
\affiliation{University of Virginia, Charlottesville, Virginia 22901, USA}
%%%end of core group
\author{S. Adhikari}
\affiliation{Florida International University, Miami, Florida 33199, USA }
\author{Z.~Akbar}
\affiliation{Florida State University, Tallahassee, Florida 32306, USA }
% ...

\date{\today}

\begin{abstract}

We measured the $g_{1}$ spin structure function of the deuteron at
low $Q^{2}$, where QCD can be approximated with chiral perturbation
theory ($\chi$PT). The data cover the resonance region, up to an
invariant mass of $W\approx1.9$~GeV. 
The generalized Gerasimov-Drell-Hearn sum, the moment $\bar{\Gamma}_{1}^{d}$
and the integral $\bar{I}_\gamma^d$ related to the
spin polarizability $\gamma_{0}^{d}$ are precisely determined down 
to a minimum $Q^2$ of 0.02~GeV$^2$ for the first time, 
about 2.5 times lower than that of previous data.
We compare them to several $\chi$PT calculations and models. These results are the
first in a program of benchmark measurements of polarization observables in the
$\chi$PT domain.

\end{abstract}

%\pacs{13.60.Hb, 11.55.Hx,25.30.Rw, 12.38.Qk}

\maketitle

For the last three decades, the spin structure of the nucleon has been actively 
studied experimentally and theoretically~\cite{Kuhn:2008sy, Aidala:2012mv}.
The reason is that spin degrees of freedom are uniquely sensitive to the details of 
the strong interaction that binds quarks into nucleons. 
The first challenge encountered by these studies was the ``spin crisis": the discovery that the 
quark spins contribute  less than expected to the proton spin~\cite{Ashman:1987hv}. 
The spin crisis brought the realization that  spin sum rules could
be used to address other challenging questions about 
quantum chromodynamics (QCD)~\cite{Anselmino:1988hn} like
quark confinement and how the low energy effective degrees of freedom of QCD (hadrons)
are related to its fundamental ones (quarks and gluons). 

This article reports the first precise measurement of the $Q^2$ evolution of the generalized Gerasimov-Drell-Hearn (GDH)
integral~\cite{Gerasimov:1965et,Helbing:2006zp} and of the 
integral $\bar{I}_\gamma^d$ related to the
spin polarizability $\gamma_0$~\cite{Guichon:1995pu} 
on the deuteron at very low four-momentum transfer $Q^2$.
Such a measurement allows us to test chiral perturbation theory ($\chi$PT)---a 
low $Q^2$ approximation of QCD---which has been challenged by earlier measurements of the
 GDH integral and of spin 
 polarizabilities~\cite{Amarian:2002ar, Slifer:2008re, Amarian:2004yf,Yun:2002td,Guler:2015hsw,Deur:2008ej,Fersch:2017}.  
 These measurements were dedicated, however,  to study QCD's hadron-parton transition.
 Only their lowest $Q^2$ points (0.05~GeV$^2$ for H and D and 0.1~GeV$^2$ for $^3$He) reached 
 the $\chi$PT domain, and with limited precision.
 The results reported here are from the Jefferson Lab (JLab) CLAS EG4 experiment, dedicated
 to measure the proton, deuteron and neutron polarized inclusive cross section at significantly 
 lower $Q^2$ than previously measured. 
A complementary program exists in JLab's Hall A, dedicated to the neutron from $^3$He~\cite{E97110}
 and to the transversely polarized proton~\cite{g2p}.

An additional goal of EG4 was to assess the reliability of extracting neutron structure information from measurements on
nuclear targets. The deuteron and $^3$He complement each other for neutron information:
nuclear binding effects in the deuteron are smaller than for $^3$He, but to obtain the neutron information
a large proton contribution is subtracted. The proton contributions in $^3$He are small, making
polarized $^3$He nearly a polarized neutron target. However, the tightly bound nucleons in $^3$He have larger 
nuclear binding effects and non-nucleonic degrees of freedom may play a larger role.

Sum rules  relate  an integral  over a dynamical quantity to a global property of the object under study. 
They offer stringent tests of the theories from which they originate. The Bjorken~\cite{Bjorken:1966jh} and 
the GDH~\cite{Gerasimov:1965et,Helbing:2006zp}  sum rules are important examples. 
The latter was originally derived for photoproduction, $Q^2=0$, and links
the helicity-dependent photoproduction cross sections $\sigma_{A}$ and $\sigma_{P}$ to the
anomalous magnetic moment $\kappa$ of the target:
\begin{equation}
\int_{\nu_0}^{\infty}\frac{\sigma_{A}(\nu)-\sigma_{P}(\nu)}{\nu}d\nu=-\frac{4\pi^2 S \alpha\kappa^2}{M^2},
\label{eq:gdh}
\end{equation}
where $M$ is the mass of the object, $S$ its spin, $\alpha$ the QED coupling, $\nu$ 
the photon energy  and $\nu_0$ the photoproduction threshold.
The $A$ and $P$ correspond to the cases where the photon spin is antiparallel and parallel to the object spin,
respectively. For the deuteron, $S=1$ and $-4\pi^2 S \alpha\kappa^2/M^2 = -0.6481(0)$ $\mu$b  
\cite{Patrignani:2016xqp}.
The GDH sum rule originates from a dispersion relation and a low energy theorem that are quite general and  
independent of QCD. The only assumption involves the convergence necessary to validate the dispersion relation.
As such, the sum rule is regarded as a solid general prediction, and experiments at MAMI, ELSA, and 
LEGS~\cite{Ahrens:2001qt} have verified it within about  7\% precision for the proton. 
Verifying the sum rule on the neutron is more difficult since no free-neutron targets exist. 
Deuteron data taken at MAMI, ELSA, and LEGS cover up to $\nu=1.8$~GeV~\cite{Ahrens:2001qt} but have not yet 
tested the sum rule due to the delicate cancellation of the deuteron photo-disintegration channel ($\approx 400~\mu$b) 
with the other inelastic channels ($\approx 401~\mu$b)~\cite{footnote}.

In the midst of the ``spin crisis," it was realized that the GDH integral could be extended to 
electroproduction to study the transition between the perturbative and 
nonperturbative domains of QCD~\cite{Anselmino:1988hn}. 
A decade later, the sum rule itself was 
generalized~\cite{Ji:1999mr, Drechsel:2002ar}:
\begin{equation}
\Gamma_1(Q^2)=\int_0^{x_0}g_1(x,Q^2)dx= \frac{Q^2}{2M^2} I_1(Q^2),
\label{eq:gdhsum_def2}
\end{equation}
where $g_1$ is the first inclusive spin structure function,  $I_1$ is the $\nu \to 0$ limit of  
the first covariant polarized VVCS amplitude, $x = Q^2/2M\nu$, and $x_0$ is the electroproduction threshold.  
The generalization connects the original GDH sum rule, Eq.~(\ref{eq:gdh}), to the 
Bjorken sum rule~\cite{Bjorken:1966jh}. 

The generalized GDH sum rule is valuable because it offers a fundamental relation for any $Q^2$.
In the low and high $Q^2$ limits where $\Gamma_1$  can be related to global properties of the target, the sum 
rule tests our understanding of the nucleon spin structure. At 
intermediate $Q^2$ it  has been used to test nonperturbative 
QCD calculations of $\Gamma_1$  such as the AdS/QCD approach~\cite{Brodsky:2010ur}, 
phenomenological models of the nucleon structure~\cite{Burkert:1992yk} and, at lower $Q^2$, $\chi$PT 
calculations~\cite{Bernard:1992nz,Ji:1999pd,Lensky:2014dda}.

An ancillary result of the present low-$Q^2$ data is their extrapolation to $Q^2=0$ 
in order to check the sum rule on $\approx$(proton+neutron)~\cite{footnote} and on the neutron. 
Although the extrapolation adds an uncertainty to these determinations, the inclusive electron scattering
used in this work sums all the reaction channels without the need to detect final state particles, 
unlike photoproduction that requires detecting each final
state, with more associated systematic uncertainties .

The GDH and Bjorken sum rules involve the first moment of the spin structure functions. 
Other sum rules exist that employ higher
moments such as the spin polarizability $\gamma_0$ sum rule, which for spin-1/2 particles reads~\cite{Drechsel:2002ar}:
\begin{eqnarray}
\label{eq:gamma_0}
\gamma_0(Q^2) = I_\gamma(Q^2) \equiv  \frac{16\alpha M^2}{Q^{6}}\int_0^{x_0}x^2\Bigl[g_1-\frac{4M^2}{Q^2}x^2g_2\Bigr]dx,
\end{eqnarray}
where $g_2$ is the second spin structure function.
An advantage of this moment is that the kinematic weighting highly suppresses the low-$x$ contribution to the sum rule, 
which typically must be estimated with model input since it is inaccessible by experiment.
For this reason, $I_\gamma$ provides a robust test of $\chi$PT, 
although it has a higher sensitivity to how data is extracted near the inelastic threshold.
$\gamma_0$ has been measured at MAMI for $Q^2=0$ and the moment $I_\gamma(Q^2)$
at JLab on the proton, neutron and deuteron for 
$0.05 \leq Q^2 \leq 4$~GeV$^2$~\cite{Yun:2002td, Amarian:2004yf, Deur:2008ej, Guler:2015hsw,Fersch:2017}.

The JLab data revealed unexpected discrepancies with $\chi$PT calculations 
for $I_\gamma(Q^2)$, its isovector and isoscalar components, as well as the generalized 
longitudinal-transverse spin polarizability $\delta_{LT}^n$~\cite{Amarian:2004yf, Yun:2002td, Deur:2008ej,Fersch:2017}.  
The data for $I_\gamma$ and $\Gamma_1$ typically agree with $\chi$PT calculations  only for the 
lowest $Q^2$ points investigated ($Q^2 \lesssim 0.07$~GeV$^2$) and generally only 
with one type of $\chi$PT calculations: for a given observable, the results of Ref.~\cite{Bernard:1992nz} 
would agree and the ones of Ref.~\cite{Ji:1999pd} would not, while the opposite occurs for another observable. 
Furthermore, the experimental and theoretical uncertainties 
of the first generation of experiments and calculations limited the usefulness of these comparisons. 
Conversely, $\Gamma_1^p - \Gamma_1^n$ was found to 
agree well with $\chi$PT~\cite{Deur:2008ej}. 
No data on $\delta_{LT}^p$ exist although some are anticipated soon~\cite{g2p}.  
This state of affairs triggered a refinement of the $\chi$PT 
calculations~\cite{Bernard:1992nz,Ji:1999pd,Lensky:2014dda} and a very low $Q^2$ 
experimental program.

The EG4 experiment took place in 2006 at JLab using the CLAS spectrometer in Hall B~\cite{Mecking:2003zu}.
The aim was to measure $g_1^p$ and $g_1^d$ over an $x$ range large enough to provide most
of the generalized GDH integral, and over a $Q^2$ range covering the region where $\chi$PT should apply. 
The inclusive scattering of polarized electrons off longitudinally polarized protons or deuterons was the 
reaction of interest, but exclusive ancillary data were also recorded~\cite{Zheng:2016ezf}.
For the deuteron run, two incident electron beam energies were used, 1.3~GeV and 2.0~GeV.
To cover the low angles necessary to reach the $Q^2$ values relevant to test $\chi$PT, 
a dedicated Cherenkov Counter (CC) was constructed and added to one of the CLAS spectrometer sectors. Furthermore, 
the target position was moved 1~m upstream of the nominal CLAS center
and the toroidal magnetic field of CLAS bent electrons outward, yielding a minimum scattering angle of about $6^o$. 
This resulted in a coverage of $0.02 \leq Q^2 \leq 0.84 $~GeV$^2$ and of invariant mass $1.07 \leq W \leq 1.9$~GeV.

The polarized beam was produced by illuminating a strained GaAs cathode with a polarized diode laser. 
A Pockels cell flipped the beam helicity pseudorandomly at  30 Hz and a half wave plate was inserted
periodically to provide an additional change of helicity sign to cancel possible false beam
asymmetries. The beam polarization varied around 85$\pm2$\% and was monitored with a M{\o}ller 
polarimeter~\cite{Mecking:2003zu}. 
The beam current ranged between 1 and 3 nA.

The polarized deuteron target consisted of  $^{15}$ND$_3$ ammonia beads held in a 1K $^4$He 
bath, and placed in a 5 T field~\cite{Crabb:1995xi}. 
The target was polarized using dynamical nuclear polarization. 
The polarization was enhanced \emph{via} irradiation with microwaves. 
The target polarization was monitored by a nuclear magnetic resonance (NMR) system and ranged between 30\% and 45\%. 
The polarization orientation was always along the beam direction.
The NMR and M{\o}ller-derived polarizations were used 
for monitoring only, the product of the beam and target polarizations for the analysis being provided 
through the measured asymmetry of quasielastic scattering. 

The scattered electrons were detected by the CLAS spectrometer. 
Besides the new CC used for data acquisition triggering
and electron identification, CLAS contained three multilayer
drift chambers that provided the momenta and charges of the
scattered particles, time-of-flight counters and 
electromagnetic calorimeters (EC) for further particle identification.
The trigger for the data acquisition system was provided by a coincidence between the new CC
and the EC. Complementary data were taken with an EC-only trigger for  efficiency measurements. 
Further information on EG4 can be found in Refs.~\cite{Zheng:2016ezf, Adhikari}.

The spin structure function $g_1$ was extracted in $W$ and $Q^2$ bins from the measured 
difference in cross sections between antiparallel and parallel beam and target polarizations:
\begin{eqnarray}
\label{eq:yield diff}
\frac{N^{\uparrow\Downarrow}(W,Q^2)}{\mathcal{L} P_b P_t  aQ_b^{\uparrow\Downarrow}} - \frac{N^{\uparrow\Uparrow}(W,Q^2)}{\mathcal{L} P_b P_t  aQ_b^{\uparrow\Uparrow}} = \Delta \sigma(W, Q^2),
\end{eqnarray}
where ``$\uparrow\Downarrow$'' or ``$\uparrow\Uparrow$" refers to beam spin 
and target polarization being antiparallel or parallel, respectively. $N$ is the number
of counts and $Q_b$ is the corresponding integrated beam charge.  $\mathcal{L}$ is a constant corresponding 
to the density of polarized target nuclei per unit area, $P_b P_t$ is the product of the beam and target polarizations and
$a(W,Q^2)$ is the detector acceptance, which also accounts for detector, trigger, and cut  efficiencies. 
$\Delta \sigma$ is the polarization dependent inclusive cross section difference in a given $(W,Q^2)$ 
bin and can be written as a linear combination of $g_1$ and $g_2$~\cite{Kuhn:2008sy, Aidala:2012mv}.
Only polarized material contributes to $\Delta \sigma$, which is  advantageous due to the dilution factor 
of the polarized targets used by EG4.

\begin{figure}[tbh!]
\includegraphics[width=8.5cm]{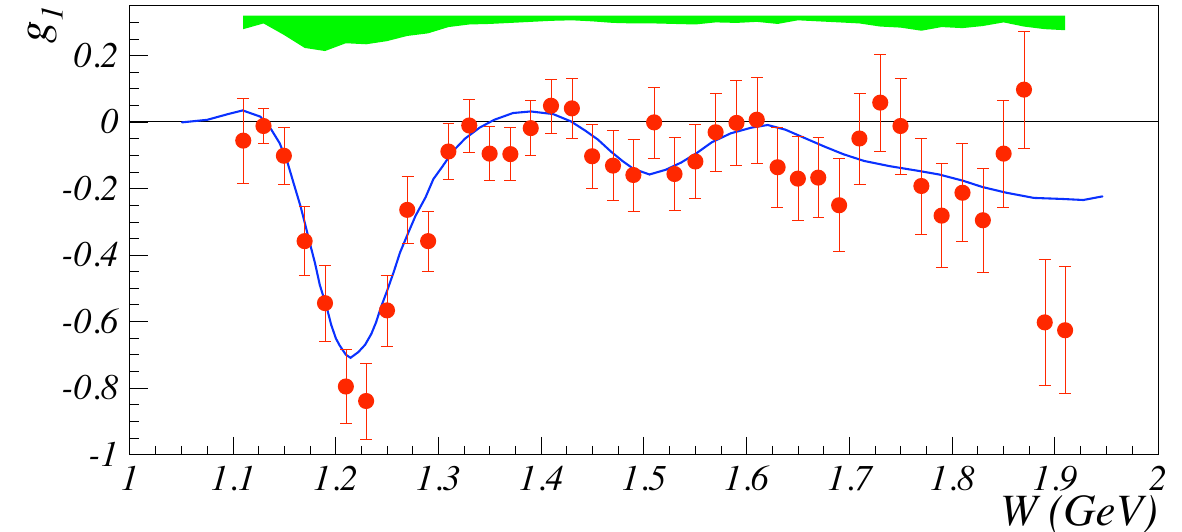}
\vspace{-0.3cm}
\caption{Example of extracted $g_1^d(W)$ vs. invariant mass $W$ (circles), together 
with the nominal value of the parameterization used for its extraction (line). The large negative peak corresponds 
to the $\Delta(1232)~3/2^+$ resonance. 
The error bars give the statistical uncertainty and the band is the total systematic uncertainty. 
The data are for $\left\langle Q^2 \right\rangle = 0.1$~GeV$^2$. 
}
\label{Fig:g1d exemple}
\end{figure}

The product of the polarized luminosity, beam and target polarizations, $P_b P_t$, and the overall electron detection 
efficiency was determined by comparing the measured yield difference in the quasielastic region, 
$0.9< W < 1$~GeV, with the calculated values. An event generator based on 
{\scriptsize RCSLACPOL}~\cite{Abe:1998wq}, with up-to-date models of structure functions and asymmetries for inelastic 
scattering from deuterium~\cite{Guler:2015hsw}, was used to generate events according to 
the fully radiated cross section. The events were followed through a full simulation of 
the CLAS spectrometer based on a G{\scriptsize EANT}-3 simulation package. 
Thus, the simulated events were analyzed in the same way as the measured data, 
thereby accounting for the bin-to-bin variation of acceptance and efficiency [Eq.~(\ref{eq:yield diff})]. 
A comparison between the simulated and the measured data in a given $Q^2$ bin is shown in Fig.~\ref{Fig:g1d exemple}.
Any deviation between the simulation and the experimental results can be due to two possible sources:
(1) A genuine difference between the $g_1$ models and the ``true'' value within that bin, and
(2) the systematic deviations of all other ingredients entering the simulation from their correct values:
this includes backgrounds and detector efficiencies and distortions, models for other structure functions 
($F_2$, $R$) and asymmetries ($A_2$), and radiative effects.
To extract $g_1(W,Q^2)$ from our measured data, we determined the amount $\delta g_1$ by which the model for 
$g_1$ had to be varied in a given bin to fully account for the difference between measured and simulated yield difference.
The systematic uncertainty on $g_1$ due to each of the sources (2) above was determined by varying one 
of the ingredients within their reasonable uncertainties and extracting the corresponding impact on $g_1$ accordingly. 
It is important to understand that although a model is used for obtaining $g_1$, there is little model dependence in the
results reported here.

Cuts were used for particle identification, to reject events not originating from the target, 
to select detector areas of high acceptance and high detector efficiency, where the detector simulation 
reproduces well the data~\cite{Adhikari}. 
Corrections were applied for contaminations from $\pi^-$ (typically less than 1\%) and from 
secondary electrons produced from photons or $\pi^0$ decay (nearly always less than 3\%).
Quality checks were performed, including detector and yield stability with time. Vertex corrections to account for the
beam raster, any target-detector misalignments and toroidal field mapping inaccuracies, were determined and applied.
Electron energy losses by ionization in the target or detector material were corrected for, as well as 
bremsstrahlung  and other radiative corrections. 
This was done using the same method as in Refs.~\cite{Yun:2002td, Fersch:2017, Guler:2015hsw}.

Systematic uncertainties are typically of the order 10\% of the extracted values for $g_1(x,Q^2)$ 
and nearly always smaller than statistical uncertainties. They are dominated by the overall 
normalization uncertainty (about 7--10\%, depending on the kinematic bin,and largely correlated), 
model uncertainties for unmeasured quantities (up to 10\% in a few kinematic bins, but normally smaller), and radiative 
corrections and kinematic uncertainties (up to 5\% near threshold but much smaller elsewhere).
These latter are mostly point-to-point uncorrelated.
The model uncertainties were estimated by modifying the parameters controlling $g_1(x,Q^2)$ and $g_2(x,Q^2)$.
The calculation and comparison of these contributions are detailed in Ref.~\cite{Adhikari}.

The complete $g_1^{d}$ data set and related moments are provided in tables as Supplemental Material.
The integrals in Eqs.~(\ref{eq:gdhsum_def2})--(\ref{eq:gamma_0}) are formed by integrating the data
over the range $x_{min}<x<x_{max}$, where $x_{max}$ corresponds to a final state mass
$W = 1.15$ GeV (assuming a target nucleon at rest) and $x_{min}$ is 
the lowest $x$ reached by the experiment for a given $Q^2$ bin. For the lowest $Q^2$ bin, 0.020~GeV$^2$, 
$x_{min}=0.0073$, and for the  largest $Q^2$ bin considered for integration, 0.592~GeV$^2$, $x_{min}=0.280$. 
The data are supplemented by the model to cover the integration range $0.001<x<x_{min}$ 
and the threshold contribution ($1.07 < W < 1.15$ GeV) at high $x$. 
There, the model is used rather than data to avoid quasielastic scattering and radiative tail contaminations~\cite{Adhikari}. 
Given the above integration range, the integrals we obtain can be considered truncated moments (in the following
indicated by over-bars) which contain only
the contributions to the full moments on the deuteron above pion production threshold and largely
exclude the two-body breakup channel. They can be approximated as
incoherent sum of the proton and neutron moments in deuterium, and hence all theory comparisons are with
the sum of the corresponding proton and neutron moments, modified by the nucleon effective polarization in the deuteron.

\begin{figure}[tb]
\includegraphics[width=8.cm]{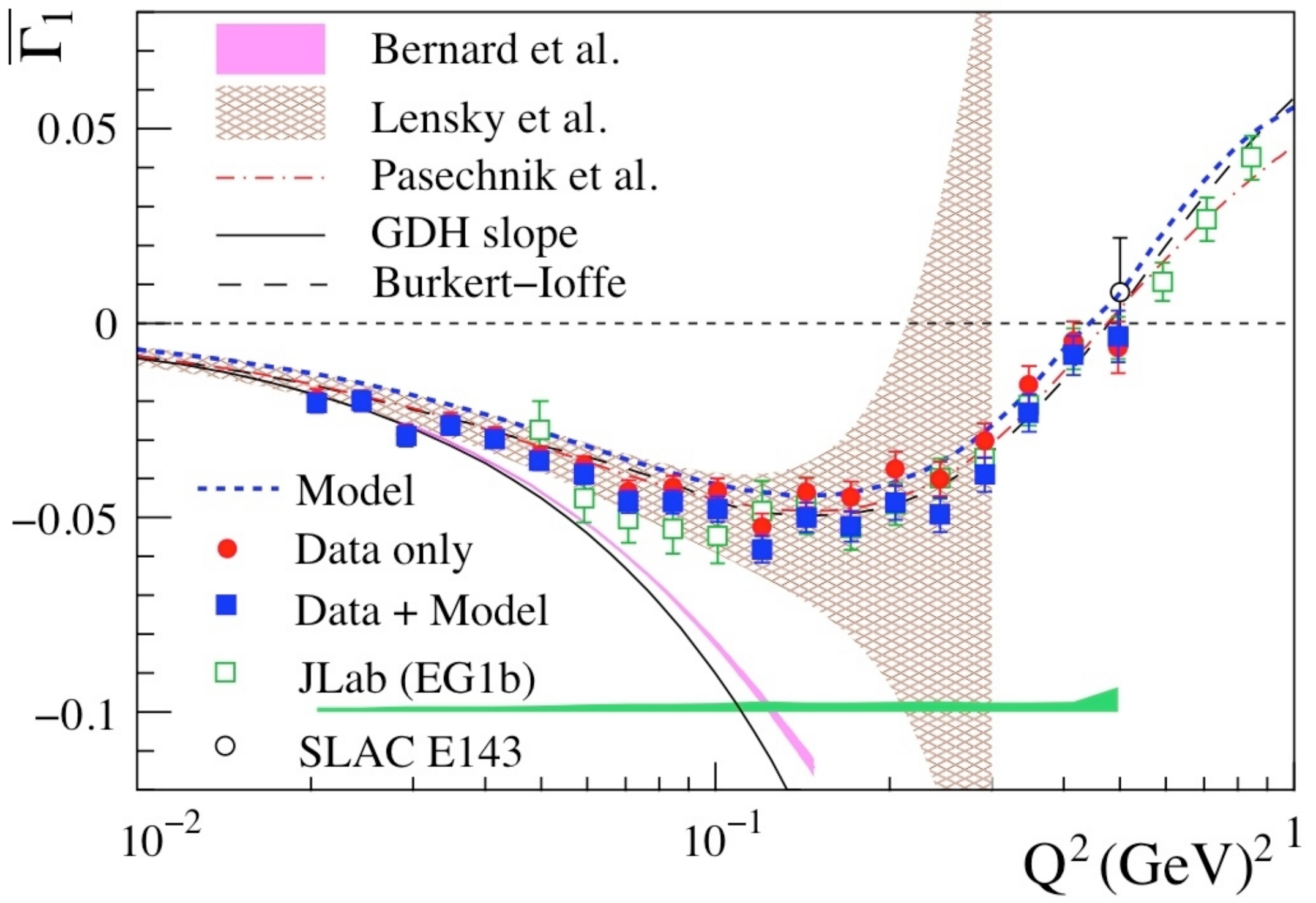}
\vspace{-0.3cm}
\caption{The truncated first moment $\bar{\Gamma}_1^d(Q^2)$.
The solid circles are the EG4 data integrated over the covered kinematics. 
The fully integrated $\bar{\Gamma}_1^d$, using a model to supplement data, is shown by the solid squares. 
The error bars are statistical. The systematic uncertainty is given by the horizontal band. The open symbols 
show data from the CLAS EG1b~\cite{Guler:2015hsw} and SLAC E143~\cite{Abe:1998wq} experiments. 
The other bands and lines show various models and $\chi$PT calculations as described in the text.
The short-dash line (Model) does not include the EG4 data, to reveal the new knowledge gained.} 
\label{Fig:GDH_D}
\end{figure}

The integral $\bar{\Gamma}_1^d(Q^2)$ is shown in Fig.~\ref{Fig:GDH_D}. The original GDH 
sum rule provides the derivative of $\Gamma_1$ at $Q^2=0$. The low-$x$ correction is small. 
The full integral (solid squares) agrees with the previous CLAS EG1b experiment~\cite{Guler:2015hsw}, 
but the minimum $Q^2$ is 2.5 times
lower. The statistical uncertainty of EG4 is improved over EG1b by about a factor of 4 at the lowest $Q^2$ points,
and thus, it allows for a more stringent test of $\chi$PT. The Lensky \emph{et al.} 
$\chi$PT calculation~\cite{Lensky:2014dda}, which supersedes the earlier calculations in Ref.~\cite{Ji:1999pd},
agrees with the data.
The most recent Bernard \emph{et al.} $\chi$PT calculation~\cite{Bernard:1992nz} 
agrees with the few  lowest $Q^2$ points.
The  Pasechnik~\emph{et al.} and Burkert-Ioffe parametrizations~\cite{Burkert:1992yk} 
describe the data well. 

The data can also be integrated to form the related moment 
$\bar{I}_{TT}^d(Q^2)$~\cite{Helbing:2006zp} extrapolated to $Q^2=0$ and compared
with the original sum rule expectation that $I_{TT}(0)=-\kappa^2/4$. 
Accounting for the deuteron $D$ state and ignoring two body breakup and coherent channels, the GDH sum rule predicts
$\bar{I}_{TT}^{d} = (1-3\omega_D/2)(I_{TT}^p+I_{TT}^n) =-1.574 \pm 0.026$, with 
$\omega_D = 0.056 \pm0.01$~\cite{omega_d}.
We extrapolated to $Q^2=0$ the data below $Q^2= 0.06$~GeV$^2$, which average at $\left\langle Q^2 \right\rangle
=0.045$~GeV$^2$. To this end, we used the (small) $Q^2$ dependence of the 
Lensky \emph{et al.} calculation~\cite{Lensky:2014dda} since it agrees very well with the data. 
We find $\bar{I}_{TT}^{d~exp}(0)=-1.724 \pm 0.027$(stat)$\pm 0.050$(syst). 
This is 10\%, or $1.5 \sigma$,  away from the sum rule prediction of $-1.574 \pm 0.026$.  
 This can be compared with the MAMI and ELSA measurement with real photons:
 $\bar{I}_{TT}^{d~exp}(0)= -1.986\pm 0.008$(stat)$\pm 0.010$(syst) integrated 
 over $0.2<\nu<1.8$~GeV (the systematic uncertainties here
 do not include any low and large $\nu$ contributions)~\cite{Ahrens:2001qt}.
 Using the proton GDH sum rule world data~\cite{Ahrens:2001qt}, we deduce the neutron GDH integral
$I_{TT}^{n~exp}(0)= -0.955\pm 0.040$(stat)$\pm 0.113$(syst), which agrees 
with the sum rule expectation $I_{TT}^{n~theo}(0)=-0.915$.

\begin{figure}[tb]
\includegraphics[width=6.3cm]{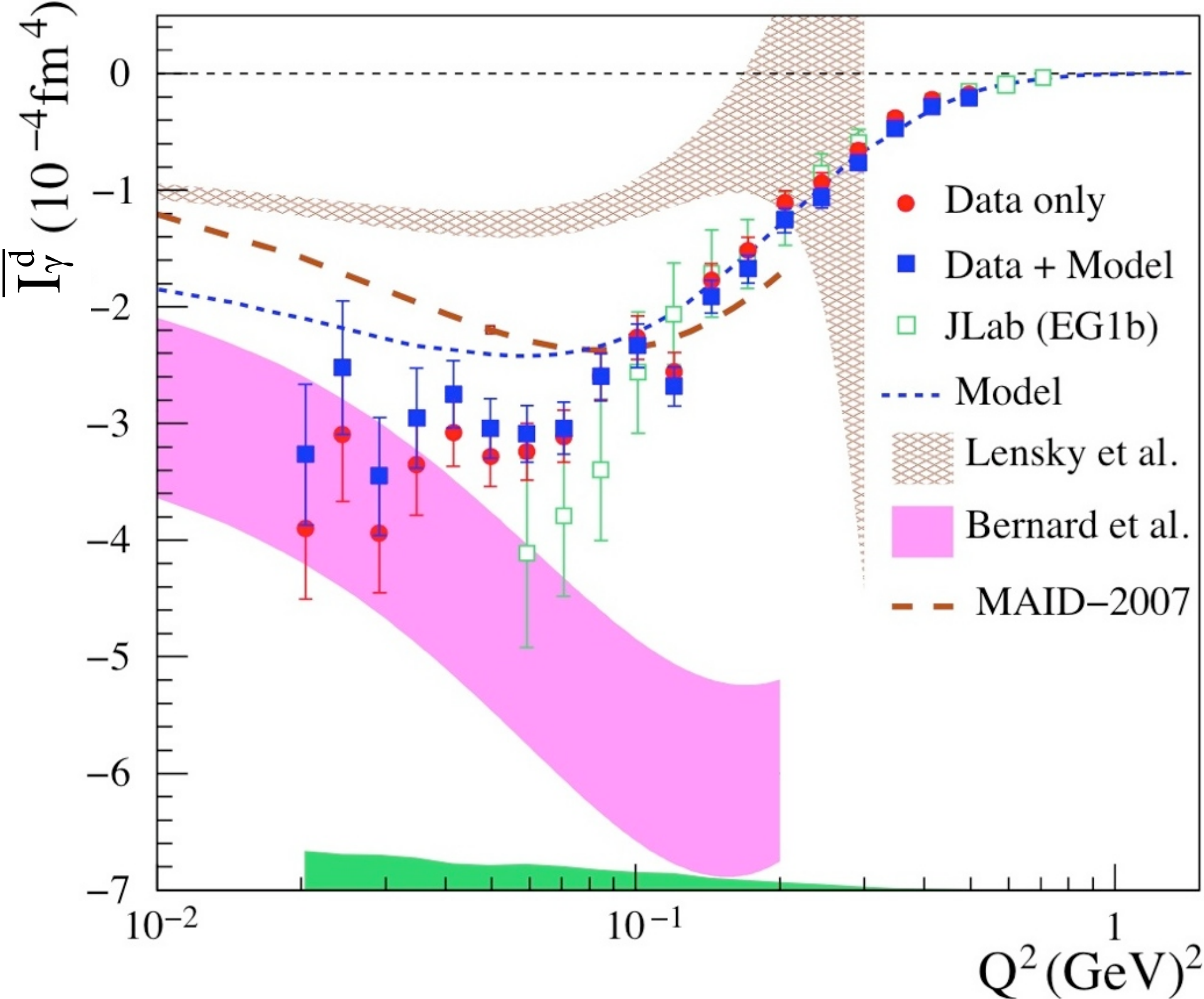}
\vspace{-0.3cm}
\caption{The truncated generalized spin polarizability integral $\bar{I}_\gamma(Q^2)$. 
See Fig.~\ref{Fig:GDH_D} for legends and theoretical calculations.
}
\label{Fig:gamma0_D}
\end{figure}

Finally, the generalized spin polarizability integral $\bar{I}_\gamma(Q^2)$ can be formed following 
Eq.~(\ref{eq:gamma_0}) and  is shown in Fig.~\ref{Fig:gamma0_D}. The MAID prediction, 
a multipole analysis of photo- and electroproduced resonance data up to $W=2$~GeV~\cite{MAID}, is relevant since the 
low-$x$ contribution, not included in MAID, is largely suppressed. The $\chi$PT calculations differ markedly. 
The full integral from EG4 (solid squares) 
agrees with the Bernard \emph{et al.} $\chi$PT calculation~\cite{Bernard:1992nz},
and it disagrees with the Lensky \emph{et al.} $\chi$PT calculation~\cite{Lensky:2014dda} 
and with the MAID model below $0.07$~GeV$^2$. 

To conclude, we report the first precise measurement of the $Q^2$ evolution of $\bar{\Gamma}_1^d$ 
and of the spin polarizability integral $\bar{I}_\gamma(Q^2)$ on the deuteron in the $0.02<Q^2<0.59$~GeV$^2$ domain.
The data reach a minimal $Q^2$ 2.5 times
lower than that of previously available data, with much improved precision. 
The degree of agreement of the different $\chi$PT methods varies
with the observable: the Bernard \emph{et al.}  calculations are more successful with $\bar{I}_\gamma$,
while the Lensky \emph{et al.} ones describe $\bar{\Gamma}_1$ well. 
Thus, no single method successfully describes both observables, and while chiral calculations are 
reaching higher precision, a satisfactory description of spin observables remains challenging. Certainly,
nuclear effects in deuterium may play a role in this.
The phenomenological models of Pasechnik~\emph{et al.} and Burkert-Ioffe agree well
with the measured $\bar{\Gamma}_1^d(Q^2)$. The MAID model disagrees with the $\bar{I}_\gamma(Q^2)$ data for $Q^2\leq0.07$~GeV$^2$.
Our data, extrapolated to $Q^2 = 0$ to check the GDH sum rule for the neutron, agree with it to within 20\%,
or about $1.0 \sigma$.

The program of providing benchmark polarization observables for $\chi$PT will be completed when the proton EG4
data become available, as well as the longitudinally and the transversally polarized data on 
the neutron ($^3$He)~\cite{E97110} and proton~\cite{g2p} from JLab's Hall A.

\acknowledgments{This work was supported by: the U.S. Department of Energy 
(DOE), the U.S. National Science Foundation, the U.S. Jeffress Memorial Trust; 
the Physics and Astronomy Department and the Office of Research and 
Economic Development at Mississippi State University, the United Kingdom's 
Science and Technology Facilities Council (STFC), the Italian Istituto Nazionale 
di Fisica Nucleare; the French Institut National de Physique Nucl\'eaire et de 
Physique des Particules, the French Centre National de la Recherche Scientifique; 
and the National Research Foundation of Korea. This material is based 
upon work supported by the U.S. Department of Energy, Office of Science, 
Office of Nuclear Physics under contract DE-AC05-06OR23177. 
}

\pagenumbering{gobble}

%\begin{sidewaystable}

\begin{table*}

\caption{CLAS EG4 experiment data table for $\Gamma_1^d$, $I_{TT}^d$ and $\gamma_0^d$.
The superscript ``full" indicates that the measured moment (``data") is complemented by Model at low-$x$.
The statistical and systematic uncertainties are the same for both ``data" and ``full".}

\normalsize{
\noindent
\begin{tabular}{|c|c|c|c|c|c|c|c|c|c|c|c|c|}
\hline 
$Q^{2}$ (GeV$^{2}$) & $\Gamma_1^{\mbox{data}}$ & $\Gamma_1^{\mbox{full}}$ & Stat. & Syst. & $I_{TT}^{\mbox{data}}$ & $I_{TT}^{\mbox{full}}$ & Stat. & Syst.& $\gamma_0^{\mbox{data}}$ & $\gamma_0^{\mbox{full}}$ & Stat. & Syst.\tabularnewline
\hline 
\hline 
 0.020 &  -0.0194 &  -0.0205 &   0.0025 &   0.0011 &  -1.9738 &  -2.0776 &   0.2489  &  0.1252 &  -3.8999 &  -3.2665 &   0.6048 &   0.3394 \tabularnewline 
 \hline
 0.024 &  -0.0199 &  -0.0201 &   0.0026 &   0.0012 &  -1.7253 &  -1.7485 &   0.2295  &  0.1151 &  -3.0963 &  -2.5206 &   0.5697 &   0.3085 \tabularnewline 
 \hline
 0.029 &  -0.0286 &  -0.0290 &   0.0027 &   0.0014 &  -2.1487 &  -2.1853 &   0.2021  &  0.1171 &  -3.9462 &  -3.4522 &   0.5043 &   0.3041 \tabularnewline 
 \hline
 0.035 &  -0.0255 &  -0.0263 &   0.0025 &   0.0015 &  -1.7120 &  -1.7651 &   0.1670  &  0.1035 &  -3.3601 &  -2.9548 &   0.4277 &   0.2799 \tabularnewline 
 \hline
 0.042 &  -0.0289 &  -0.0300 &   0.0021 &   0.0015 &  -1.6658 &  -1.7296 &   0.1212  &  0.0872 &  -3.0780 &  -2.7518 &   0.2878 &   0.2322 \tabularnewline 
 \hline
 0.050 &  -0.0340 &  -0.0355 &   0.0023 &   0.0017 &  -1.7351 &  -1.8125 &   0.1119  &  0.0849 &  -3.2859 &  -3.0435 &   0.2565 &   0.2191 \tabularnewline 
 \hline
 0.059 &  -0.0367 &  -0.0389 &   0.0025 &   0.0019 &  -1.6804 &  -1.7729 &   0.1087  &  0.0880 &  -3.2418 &  -3.0891 &   0.2441 &   0.2266 \tabularnewline
\hline
\end{tabular}
}
%\end{sidewaystable}
\end{table*}
\clearpage

\begin{table}
%scriptsize{
\vspace{2cm}
Table II. CLAS EG4 experiment data table for $g_1$ and $A_1/F1$ on the deuteron.

~

\begin{tabular}{|c|c|c|c|c|c|c|c|}
\hline 
$Q^{2}$ (GeV$^{2}$) & W (GeV) & $g_{1}$ & Stat. & Syst. & $A_{1}/F_{1}$ & Stat. & Syst.\tabularnewline
\hline 
\hline 
 0.017 &     1.13 &    0.353 &    0.413 &    0.060 &    0.436 &    0.558 &    0.077 \tabularnewline 
 \hline
 0.017 &     1.15 &    0.760 &    0.362 &    0.102 &    0.936 &    0.464 &    0.129 \tabularnewline 
 \hline
 0.017 &     1.17 &   -0.536 &    0.348 &    0.170 &   -0.692 &    0.430 &    0.214 \tabularnewline 
 \hline
 0.017 &     1.19 &   -0.400 &    0.353 &    0.160 &   -0.496 &    0.422 &    0.208 \tabularnewline 
 \hline
 0.017 &     1.21 &   -0.900 &    0.369 &    0.174 &   -1.060 &    0.431 &    0.221 \tabularnewline 
 \hline
 0.017 &     1.23 &   -1.209 &    0.349 &    0.183 &   -1.391 &    0.398 &    0.220 \tabularnewline 
 \hline
 0.017 &     1.25 &   -0.691 &    0.337 &    0.133 &   -0.790 &    0.378 &    0.155 \tabularnewline
\hline
 \end{tabular}
%\caption{CLAS EG4 experiment data table for $g_1$ and $A_1/F1$ on the deuteron.}
\end{table}
%\clearpage
\begin{table}
%scriptsize{
\vspace{2.3cm}
\begin{tabular}{|c|c|c|c|c|c|c|c|}
 \hline
$Q^{2}$ (GeV$^{2}$) & W (GeV) & $g_{1}$ & Stat. & Syst. & $A_{1}/F_{1}$ & Stat. & Syst.\tabularnewline
\hline 
 0.020 &     1.29 &   -0.242 &    0.278 &    0.080 &   -0.288 &    0.308 &    0.088 \tabularnewline 
 \hline
 0.020 &     1.31 &    0.179 &    0.267 &    0.055 &    0.177 &    0.293 &    0.056 \tabularnewline 
 \hline
 0.020 &     1.33 &   -0.298 &    0.277 &    0.046 &   -0.343 &    0.301 &    0.049 \tabularnewline 
 \hline
 0.020 &     1.35 &   -0.190 &    0.292 &    0.033 &   -0.224 &    0.314 &    0.035 \tabularnewline 
 \hline
 0.020 &     1.37 &   -0.140 &    0.323 &    0.023 &   -0.169 &    0.345 &    0.025 \tabularnewline 
 \hline
 0.020 &     1.39 &   -0.150 &    0.366 &    0.029 &   -0.180 &    0.389 &    0.034 \tabularnewline 
 \hline
 0.020 &     1.41 &   -0.443 &    0.360 &    0.045 &   -0.490 &    0.380 &    0.049 \tabularnewline
\hline
  \end{tabular}
\end{table}
\clearpage
\begin{table}
%scriptsize{
\vspace{2cm}
\begin{tabular}{|c|c|c|c|c|c|c|c|}
 \hline
$Q^{2}$ (GeV$^{2}$) & W (GeV) & $g_{1}$ & Stat. & Syst. & $A_{1}/F_{1}$ & Stat. & Syst.\tabularnewline
\hline 
 0.024 &     1.39 &   -0.215 &    0.275 &    0.024 &   -0.256 &    0.295 &    0.025 \tabularnewline 
 \hline
 0.024 &     1.41 &   -0.018 &    0.327 &    0.026 &   -0.045 &    0.348 &    0.027 \tabularnewline 
 \hline
 0.024 &     1.43 &   -0.876 &    0.287 &    0.081 &   -0.956 &    0.304 &    0.084 \tabularnewline 
 \hline
 0.024 &     1.45 &   -0.950 &    0.341 &    0.076 &   -1.029 &    0.360 &    0.080 \tabularnewline 
 \hline
 0.024 &     1.47 &   -0.369 &    0.360 &    0.024 &   -0.414 &    0.378 &    0.026 \tabularnewline 
 \hline
 0.024 &     1.49 &   -0.119 &    0.361 &    0.016 &   -0.150 &    0.377 &    0.021 \tabularnewline 
 \hline
 0.024 &     1.51 &   -1.586 &    0.406 &    0.130 &   -1.675 &    0.423 &    0.136 \tabularnewline
\hline
  \end{tabular}
\end{table}
%\clearpage
\begin{table}
%scriptsize{
\vspace{2cm}
\begin{tabular}{|c|c|c|c|c|c|c|c|} \hline
$Q^{2}$ (GeV$^{2}$) & W (GeV) & $g_{1}$ & Stat. & Syst. & $A_{1}/F_{1}$ & Stat. & Syst.\tabularnewline
 \hline
 0.029 &     1.49 &   -0.819 &    0.205 &    0.065 &   -0.895 &    0.216 &    0.068 \tabularnewline 
 \hline
 0.029 &     1.51 &   -0.614 &    0.192 &    0.050 &   -0.672 &    0.201 &    0.052 \tabularnewline 
 \hline
 0.029 &     1.53 &   -0.185 &    0.177 &    0.021 &   -0.220 &    0.186 &    0.021 \tabularnewline 
 \hline
 0.029 &     1.55 &   -0.264 &    0.178 &    0.029 &   -0.301 &    0.185 &    0.029 \tabularnewline 
 \hline
 0.029 &     1.57 &   -0.294 &    0.179 &    0.036 &   -0.329 &    0.186 &    0.036 \tabularnewline 
 \hline
 0.029 &     1.59 &   -0.511 &    0.175 &    0.051 &   -0.550 &    0.181 &    0.052 \tabularnewline 
 \hline
 0.029 &     1.61 &   -0.460 &    0.221 &    0.036 &   -0.494 &    0.228 &    0.038 \tabularnewline
\hline
  \end{tabular}
\end{table}\clearpage
\begin{table}
%scriptsize{
\vspace{2cm}
\begin{tabular}{|c|c|c|c|c|c|c|c|} \hline
$Q^{2}$ (GeV$^{2}$) & W (GeV) & $g_{1}$ & Stat. & Syst. & $A_{1}/F_{1}$ & Stat. & Syst.\tabularnewline
 \hline
 0.035 &     1.59 &   -0.394 &    0.170 &    0.036 &   -0.437 &    0.177 &    0.036 \tabularnewline 
 \hline
 0.035 &     1.61 &    0.015 &    0.194 &    0.018 &   -0.006 &    0.202 &    0.019 \tabularnewline 
 \hline
 0.035 &     1.63 &   -0.381 &    0.194 &    0.026 &   -0.412 &    0.202 &    0.027 \tabularnewline 
 \hline
 0.035 &     1.65 &   -0.261 &    0.182 &    0.015 &   -0.284 &    0.188 &    0.015 \tabularnewline 
 \hline
 0.035 &     1.67 &   -0.373 &    0.202 &    0.023 &   -0.395 &    0.209 &    0.024 \tabularnewline 
 \hline
 0.035 &     1.69 &   -0.647 &    0.205 &    0.047 &   -0.674 &    0.211 &    0.048 \tabularnewline 
 \hline
 0.035 &     1.71 &   -0.260 &    0.198 &    0.028 &   -0.272 &    0.203 &    0.027 \tabularnewline
\hline
  \end{tabular}
\end{table}
%\clearpage
\begin{table}
%scriptsize{
\vspace{2cm}
\begin{tabular}{|c|c|c|c|c|c|c|c|} \hline
$Q^{2}$ (GeV$^{2}$) & W (GeV) & $g_{1}$ & Stat. & Syst. & $A_{1}/F_{1}$ & Stat. & Syst.\tabularnewline
 \hline
 0.042 &     1.67 &   -0.902 &    0.192 &    0.073 &   -0.949 &    0.199 &    0.076 \tabularnewline 
 \hline
 0.042 &     1.69 &   -0.478 &    0.175 &    0.034 &   -0.503 &    0.181 &    0.035 \tabularnewline 
 \hline
 0.042 &     1.71 &   -0.171 &    0.186 &    0.028 &   -0.183 &    0.192 &    0.030 \tabularnewline 
 \hline
 0.042 &     1.73 &   -0.642 &    0.198 &    0.065 &   -0.666 &    0.204 &    0.067 \tabularnewline 
 \hline
 0.042 &     1.75 &   -0.115 &    0.200 &    0.052 &   -0.122 &    0.205 &    0.054 \tabularnewline 
 \hline
 0.042 &     1.77 &   -0.486 &    0.232 &    0.059 &   -0.504 &    0.238 &    0.061 \tabularnewline 
 \hline
 0.042 &     1.79 &   -0.365 &    0.238 &    0.048 &   -0.380 &    0.244 &    0.049 \tabularnewline
\hline
  \end{tabular}
\end{table}\clearpage
\begin{table}
%scriptsize{
\vspace{2cm}
\begin{tabular}{|c|c|c|c|c|c|c|c|} \hline
$Q^{2}$ (GeV$^{2}$) & W (GeV) & $g_{1}$ & Stat. & Syst. & $A_{1}/F_{1}$ & Stat. & Syst.\tabularnewline
 \hline
 0.050 &     1.75 &   -0.536 &    0.193 &    0.060 &   -0.559 &    0.200 &    0.062 \tabularnewline 
 \hline
 0.050 &     1.77 &   -0.031 &    0.194 &    0.062 &   -0.037 &    0.200 &    0.064 \tabularnewline 
 \hline
 0.050 &     1.79 &   -0.282 &    0.195 &    0.049 &   -0.297 &    0.201 &    0.050 \tabularnewline 
 \hline
 0.050 &     1.81 &   -0.294 &    0.220 &    0.030 &   -0.308 &    0.226 &    0.030 \tabularnewline 
 \hline
 0.050 &     1.83 &   -0.085 &    0.230 &    0.029 &   -0.093 &    0.236 &    0.030 \tabularnewline 
 \hline
 0.050 &     1.85 &   -0.287 &    0.223 &    0.034 &   -0.299 &    0.229 &    0.036 \tabularnewline 
 \hline
 0.050 &     1.87 &   -0.648 &    0.270 &    0.041 &   -0.667 &    0.276 &    0.043 \tabularnewline
\hline
  \end{tabular}
\end{table}
%\clearpage
\begin{table}
%scriptsize{
\vspace{1.8cm}
\begin{tabular}{|c|c|c|c|c|c|c|c|} \hline
$Q^{2}$ (GeV$^{2}$) & W (GeV) & $g_{1}$ & Stat. & Syst. & $A_{1}/F_{1}$ & Stat. & Syst.\tabularnewline
 \hline
 0.059 &     1.83 &   -0.123 &    0.207 &    0.023 &   -0.134 &    0.214 &    0.023 \tabularnewline 
 \hline
 0.059 &     1.85 &   -0.130 &    0.202 &    0.023 &   -0.139 &    0.208 &    0.023 \tabularnewline 
 \hline
 0.059 &     1.87 &   -0.262 &    0.230 &    0.016 &   -0.273 &    0.237 &    0.017 \tabularnewline 
 \hline
 0.059 &     1.89 &   -0.664 &    0.235 &    0.042 &   -0.684 &    0.242 &    0.043 \tabularnewline 
 \hline
 0.059 &     1.91 &   -0.271 &    0.274 &    0.010 &   -0.278 &    0.281 &    0.008 \tabularnewline 
 \hline
 0.071 &     1.11 &   -0.007 &    0.036 &    0.018 &   -0.080 &    0.086 &    0.034 \tabularnewline 
 \hline
 0.071 &     1.13 &   -0.061 &    0.095 &    0.027 &   -0.214 &    0.202 &    0.058 \tabularnewline
\hline
  \end{tabular}
\end{table}\clearpage
\begin{table}
%scriptsize{
\vspace{2cm}
\begin{tabular}{|c|c|c|c|c|c|c|c|} \hline
$Q^{2}$ (GeV$^{2}$) & W (GeV) & $g_{1}$ & Stat. & Syst. & $A_{1}/F_{1}$ & Stat. & Syst.\tabularnewline
 \hline
 0.084 &     1.11 &    0.009 &    0.078 &    0.025 &   -0.042 &    0.199 &    0.062 \tabularnewline 
 \hline
 0.084 &     1.13 &   -0.008 &    0.090 &    0.031 &   -0.104 &    0.206 &    0.072 \tabularnewline 
 \hline
 0.084 &     1.15 &   -0.142 &    0.089 &    0.063 &   -0.381 &    0.184 &    0.129 \tabularnewline 
 \hline
 0.084 &     1.17 &   -0.443 &    0.103 &    0.099 &   -0.904 &    0.195 &    0.186 \tabularnewline 
 \hline
 0.084 &     1.19 &   -0.802 &    0.118 &    0.119 &   -1.447 &    0.208 &    0.205 \tabularnewline 
 \hline
 0.084 &     1.21 &   -0.729 &    0.122 &    0.081 &   -1.226 &    0.203 &    0.126 \tabularnewline 
 \hline
 0.084 &     1.23 &   -0.768 &    0.117 &    0.079 &   -1.233 &    0.185 &    0.118 \tabularnewline
\hline
  \end{tabular}
\end{table}
%\clearpage
\begin{table}
%scriptsize{
\vspace{2cm}
\begin{tabular}{|c|c|c|c|c|c|c|c|} \hline
$Q^{2}$ (GeV$^{2}$) & W (GeV) & $g_{1}$ & Stat. & Syst. & $A_{1}/F_{1}$ & Stat. & Syst.\tabularnewline
 \hline
 0.101 &     1.19 &   -0.545 &    0.115 &    0.106 &   -1.044 &    0.215 &    0.192 \tabularnewline 
 \hline
 0.101 &     1.21 &   -0.796 &    0.111 &    0.083 &   -1.403 &    0.194 &    0.135 \tabularnewline 
 \hline
 0.101 &     1.23 &   -0.839 &    0.114 &    0.086 &   -1.409 &    0.189 &    0.132 \tabularnewline 
 \hline
 0.101 &     1.25 &   -0.567 &    0.107 &    0.078 &   -0.936 &    0.168 &    0.116 \tabularnewline 
 \hline
 0.101 &     1.27 &   -0.264 &    0.101 &    0.062 &   -0.462 &    0.151 &    0.089 \tabularnewline 
 \hline
 0.101 &     1.29 &   -0.358 &    0.091 &    0.053 &   -0.594 &    0.131 &    0.073 \tabularnewline 
 \hline
 0.101 &     1.31 &   -0.088 &    0.086 &    0.035 &   -0.207 &    0.121 &    0.045 \tabularnewline
\hline
  \end{tabular}
\end{table}\clearpage
\begin{table}
%scriptsize{
\vspace{2cm}
\begin{tabular}{|c|c|c|c|c|c|c|c|} \hline
$Q^{2}$ (GeV$^{2}$) & W (GeV) & $g_{1}$ & Stat. & Syst. & $A_{1}/F_{1}$ & Stat. & Syst.\tabularnewline
 \hline
 0.120 &     1.27 &   -0.344 &    0.094 &    0.065 &   -0.608 &    0.149 &    0.097 \tabularnewline 
 \hline
 0.120 &     1.29 &   -0.251 &    0.085 &    0.048 &   -0.462 &    0.129 &    0.068 \tabularnewline 
 \hline
 0.120 &     1.31 &   -0.005 &    0.082 &    0.031 &   -0.098 &    0.121 &    0.041 \tabularnewline 
 \hline
 0.120 &     1.33 &   -0.032 &    0.082 &    0.024 &   -0.144 &    0.116 &    0.031 \tabularnewline 
 \hline
 0.120 &     1.35 &    0.093 &    0.076 &    0.021 &    0.024 &    0.104 &    0.025 \tabularnewline 
 \hline
 0.120 &     1.37 &    0.128 &    0.082 &    0.019 &    0.062 &    0.110 &    0.021 \tabularnewline 
 \hline
 0.120 &     1.39 &   -0.033 &    0.078 &    0.023 &   -0.157 &    0.101 &    0.025 \tabularnewline
\hline
  \end{tabular}
\end{table}
%\clearpage
\begin{table}
%scriptsize{
\vspace{2cm}
\begin{tabular}{|c|c|c|c|c|c|c|c|} \hline
$Q^{2}$ (GeV$^{2}$) & W (GeV) & $g_{1}$ & Stat. & Syst. & $A_{1}/F_{1}$ & Stat. & Syst.\tabularnewline
 \hline
 0.144 &     1.39 &   -0.035 &    0.075 &    0.017 &   -0.172 &    0.101 &    0.019 \tabularnewline 
 \hline
 0.144 &     1.41 &   -0.026 &    0.081 &    0.017 &   -0.164 &    0.107 &    0.019 \tabularnewline 
 \hline
 0.144 &     1.43 &    0.046 &    0.085 &    0.014 &   -0.073 &    0.110 &    0.015 \tabularnewline 
 \hline
 0.144 &     1.45 &   -0.117 &    0.096 &    0.018 &   -0.280 &    0.121 &    0.020 \tabularnewline 
 \hline
 0.144 &     1.47 &   -0.006 &    0.108 &    0.019 &   -0.137 &    0.135 &    0.020 \tabularnewline 
 \hline
 0.144 &     1.49 &    0.006 &    0.110 &    0.021 &   -0.113 &    0.134 &    0.022 \tabularnewline 
 \hline
 0.144 &     1.51 &    0.066 &    0.103 &    0.025 &   -0.030 &    0.124 &    0.026 \tabularnewline
\hline
  \end{tabular}
\end{table}\clearpage
\begin{table}
%scriptsize{
\vspace{2cm}
\begin{tabular}{|c|c|c|c|c|c|c|c|} \hline
$Q^{2}$ (GeV$^{2}$) & W (GeV) & $g_{1}$ & Stat. & Syst. & $A_{1}/F_{1}$ & Stat. & Syst.\tabularnewline
 \hline
 0.171 &     1.47 &    0.148 &    0.084 &    0.023 &    0.046 &    0.108 &    0.025 \tabularnewline 
 \hline
 0.171 &     1.49 &    0.062 &    0.089 &    0.022 &   -0.057 &    0.113 &    0.023 \tabularnewline 
 \hline
 0.171 &     1.51 &    0.209 &    0.095 &    0.034 &    0.137 &    0.119 &    0.037 \tabularnewline 
 \hline
 0.171 &     1.53 &    0.045 &    0.093 &    0.030 &   -0.059 &    0.113 &    0.032 \tabularnewline 
 \hline
 0.171 &     1.55 &    0.031 &    0.115 &    0.032 &   -0.072 &    0.139 &    0.049 \tabularnewline 
 \hline
 0.171 &     1.57 &    0.215 &    0.113 &    0.043 &    0.153 &    0.135 &    0.047 \tabularnewline 
 \hline
 0.171 &     1.59 &    0.122 &    0.109 &    0.035 &    0.047 &    0.129 &    0.043 \tabularnewline
\hline
  \end{tabular}
\end{table}
%\clearpage
\begin{table}
%scriptsize{
\vspace{2cm}
\begin{tabular}{|c|c|c|c|c|c|c|c|} \hline
$Q^{2}$ (GeV$^{2}$) & W (GeV) & $g_{1}$ & Stat. & Syst. & $A_{1}/F_{1}$ & Stat. & Syst.\tabularnewline
 \hline
 0.205 &     1.57 &    0.101 &    0.106 &    0.040 &    0.009 &    0.129 &    0.048 \tabularnewline 
 \hline
 0.205 &     1.59 &    0.167 &    0.100 &    0.044 &    0.095 &    0.121 &    0.047 \tabularnewline 
 \hline
 0.205 &     1.61 &   -0.142 &    0.114 &    0.033 &   -0.267 &    0.136 &    0.038 \tabularnewline 
 \hline
 0.205 &     1.63 &    0.006 &    0.101 &    0.030 &   -0.079 &    0.120 &    0.035 \tabularnewline 
 \hline
 0.205 &     1.65 &    0.133 &    0.100 &    0.034 &    0.083 &    0.116 &    0.033 \tabularnewline 
 \hline
 0.205 &     1.67 &   -0.183 &    0.115 &    0.022 &   -0.273 &    0.133 &    0.034 \tabularnewline 
 \hline
 0.205 &     1.69 &   -0.024 &    0.120 &    0.013 &   -0.080 &    0.138 &    0.034 \tabularnewline
\hline
  \end{tabular}
\end{table}\clearpage
\begin{table}
%scriptsize{
\vspace{2cm}
\begin{tabular}{|c|c|c|c|c|c|c|c|} \hline
$Q^{2}$ (GeV$^{2}$) & W (GeV) & $g_{1}$ & Stat. & Syst. & $A_{1}/F_{1}$ & Stat. & Syst.\tabularnewline
 \hline
 0.244 &     1.71 &    0.050 &    0.101 &    0.034 &    0.008 &    0.118 &    0.033 \tabularnewline 
 \hline
 0.244 &     1.73 &   -0.020 &    0.108 &    0.036 &   -0.065 &    0.125 &    0.039 \tabularnewline 
 \hline
 0.244 &     1.75 &    0.107 &    0.125 &    0.032 &    0.086 &    0.143 &    0.035 \tabularnewline 
 \hline
 0.244 &     1.77 &   -0.032 &    0.140 &    0.025 &   -0.069 &    0.159 &    0.030 \tabularnewline 
 \hline
 0.244 &     1.79 &   -0.051 &    0.231 &    0.028 &   -0.087 &    0.261 &    0.031 \tabularnewline 
 \hline
 0.244 &     1.81 &   -0.380 &    0.248 &    0.047 &   -0.452 &    0.278 &    0.054 \tabularnewline 
 \hline
 0.244 &     1.83 &    0.012 &    0.188 &    0.042 &   -0.010 &    0.209 &    0.048 \tabularnewline
\hline
  \end{tabular}
\end{table}
%\clearpage
\begin{table}
%scriptsize{
\vspace{2cm}
\begin{tabular}{|c|c|c|c|c|c|c|c|} \hline
$Q^{2}$ (GeV$^{2}$) & W (GeV) & $g_{1}$ & Stat. & Syst. & $A_{1}/F_{1}$ & Stat. & Syst.\tabularnewline
 \hline
 0.348 &     1.17 &   -0.209 &    0.069 &    0.036 &   -0.560 &    0.190 &    0.097 \tabularnewline 
 \hline
 0.348 &     1.19 &   -0.235 &    0.078 &    0.032 &   -0.565 &    0.200 &    0.076 \tabularnewline 
 \hline
 0.348 &     1.21 &   -0.295 &    0.077 &    0.030 &   -0.667 &    0.187 &    0.064 \tabularnewline 
 \hline
 0.348 &     1.23 &   -0.444 &    0.084 &    0.043 &   -0.980 &    0.191 &    0.088 \tabularnewline 
 \hline
 0.348 &     1.25 &   -0.248 &    0.078 &    0.036 &   -0.538 &    0.167 &    0.068 \tabularnewline 
 \hline
 0.348 &     1.27 &   -0.191 &    0.071 &    0.034 &   -0.427 &    0.146 &    0.065 \tabularnewline 
 \hline
 0.348 &     1.29 &   -0.069 &    0.070 &    0.030 &   -0.200 &    0.137 &    0.055 \tabularnewline
\hline
  \end{tabular}
\end{table}\clearpage
\begin{table}
%scriptsize{
\vspace{2cm}
\begin{tabular}{|c|c|c|c|c|c|c|c|} \hline
$Q^{2}$ (GeV$^{2}$) & W (GeV) & $g_{1}$ & Stat. & Syst. & $A_{1}/F_{1}$ & Stat. & Syst.\tabularnewline
 \hline
 0.416 &     1.39 &    0.099 &    0.059 &    0.031 &    0.015 &    0.099 &    0.032 \tabularnewline 
 \hline
 0.416 &     1.41 &    0.048 &    0.082 &    0.034 &   -0.083 &    0.133 &    0.043 \tabularnewline 
 \hline
 0.416 &     1.43 &    0.091 &    0.082 &    0.054 &   -0.023 &    0.129 &    0.041 \tabularnewline 
 \hline
 0.416 &     1.45 &    0.059 &    0.106 &    0.049 &   -0.080 &    0.164 &    0.043 \tabularnewline 
 \hline
 0.416 &     1.47 &    0.204 &    0.082 &    0.067 &    0.134 &    0.124 &    0.050 \tabularnewline 
 \hline
 0.416 &     1.49 &    0.146 &    0.124 &    0.055 &    0.047 &    0.183 &    0.083 \tabularnewline 
 \hline
 0.416 &     1.51 &    0.306 &    0.115 &    0.077 &    0.278 &    0.166 &    0.092 \tabularnewline
\hline
  \end{tabular}
\end{table}
%\clearpage
\begin{table}
%scriptsize{
\vspace{2cm}
\begin{tabular}{|c|c|c|c|c|c|c|c|} \hline
$Q^{2}$ (GeV$^{2}$) & W (GeV) & $g_{1}$ & Stat. & Syst. & $A_{1}/F_{1}$ & Stat. & Syst.\tabularnewline
 \hline
 0.592 &     1.11 &   -0.026 &    0.014 &    0.005 &   -0.092 &    0.048 &    0.017 \tabularnewline 
 \hline
 0.592 &     1.13 &   -0.098 &    0.063 &    0.014 &   -0.309 &    0.199 &    0.045 \tabularnewline 
 \hline
 0.592 &     1.15 &   -0.117 &    0.079 &    0.013 &   -0.336 &    0.232 &    0.055 \tabularnewline 
 \hline
 0.592 &     1.17 &   -0.155 &    0.064 &    0.021 &   -0.410 &    0.178 &    0.063 \tabularnewline 
 \hline
 0.592 &     1.19 &   -0.197 &    0.070 &    0.025 &   -0.486 &    0.185 &    0.068 \tabularnewline 
 \hline
 0.592 &     1.21 &   -0.312 &    0.080 &    0.039 &   -0.747 &    0.202 &    0.091 \tabularnewline 
 \hline
 0.592 &     1.23 &   -0.171 &    0.071 &    0.023 &   -0.384 &    0.168 &    0.055 \tabularnewline
\hline
\end{tabular}
\end{table}\clearpage


\begin{thebibliography}{00}
\bibitem{footnote}In this Letter, we exclude most of the deuteron break-up channel by requiring 
$W = \sqrt{M^2 + 2 M \nu - Q^2} > 1.15$ GeV. In that case, the integrals are close to the incoherent sums 
of the corresponding integrals for the proton and the neutron, modified by the deuteron $D$ state correction factor.
\end{thebibliography}
\end{document}